# Jupiter's influence on the building blocks of Mars and Earth

R. Brasser[1†], N. Dauphas[2] and S. J. Mojzsis[3,4†]

[1]Earth Life Science Institute, Tokyo Institute of Technology, Ookayama, Meguro-ku, Tokyo 152-8550, Japan

[2]Origins Laboratory, Department of the Geophysical Sciences and Enrico Fermi Institute, The University of Chicago, 5734 South Ellis Avenue, Chicago, Illinois 60637, USA

[3]Department of Geological Sciences, University of Colorado, UCB 399, 2200 Colorado Avenue, Boulder, Colorado 80309-0399, USA

[4]Institute for Geological and Geochemical Research, Research Centre for Astronomy and Earth Sciences, Hungarian Academy of Sciences, 45 Budaörsi Street, H-1112 Budapest, Hungary

[†]Collaborative for Research in Origins (CRiO), The John Templeton Foundation – FfAME Origins Program

Editorial correspondence:

Ramon Brasser      email: brasser_astro@yahoo.com (t: +81 3 5734 3414; f: +81 3 5734 3416)

Stephen J. Mojzsis      email: mojzsis@colorado.edu (t: +1 303 492 5015; f: +1 303 492 2606)

**Abstract.**
Radiometric dating indicates that Mars accreted in the first ~4 Myr of solar system formation, which coincides with the formation and possible migration of Jupiter. While nebular gas from the protoplanetary disk was still present, Jupiter may have migrated inwards and tacked at 1.5 AU in a 3:2 resonance with Saturn. This migration excited planetary building blocks in the inner solar system, resulting in extensive mixing and planetesimal removal. Here we evaluate the plausible nature of Mars' building blocks, focusing in particular on how its growth was influenced by the formation and migration of Jupiter. We use a combination of dynamical simulations and an isotopic mixing model that traces the accretion of elements with different affinities for metal. Dynamical simulations show that Jupiter's migration causes the late stages of Earth's and Mars' accretion to be dominated by EC-type (enstatite chondrite) material due to the loss of OC (ordinary chondrite) planetesimals. Our analysis of available isotopic data for SNC meteorites shows that Mars consists of approximately $68\%^{+0}_{-39}$ EC+$32\%^{+35}_{-0}$ OC by mass ($2\sigma$). The large uncertainties indicate that isotopic analyses of martian samples are for the most part too imprecise to definitely test model predictions; in particular it remains uncertain whether or not Mars accreted predominantly EC material in the latter stages of its formation history. Dynamical simulations also provide no definitive constraint on Mars' accretion history due to the great variety of dynamical pathways that the martian embryo exhibits. The present work calls for new measurements of isotopic anomalies in SNC meteorites targeting siderophile elements (most notably Ni, Mo and Ru) to constrain Mars' accretion history and its formation location.

# 1. Introduction

Mars formed rapidly, within about 4 Myr after solar system formation [Dauphas & Pourmand, 2011]. This is within the lifetime of the protoplanetary nebula. Kruijer et al. [2017] inferred from a dichotomy in Mo isotopic anomalies among iron meteorites that Jupiter's core may have formed within 1 Myr of solar system formation. Furthermore, Jupiter must have completed its growth before the nebular gas had dissipated, which studies of remnant magnetism in primitive meteorites [Wang et al., 2016] and observations of young stellar objects [Mamajek, 2009] indicate takes place within the first ~5 Myr of solar system formation. If Jupiter underwent gas-driven migration, as has been proposed in the Grand Tack model [Walsh et al. 2011], then this migration presumably occurred while Mars was still growing.

Two recent studies point to two very different scenarios for Mars' accretion. Using a Monte Carlo mixing model for a subset of isotopic systems, Fitoussi et al. [2016] report that Earth and Mars are built from up to 93% of the same material, and that a scenario wherein Mars forms beyond 2 AU is inconsistent with mixing calculations. Even though dynamical simulations were not reported in that work, a preferred formation model was advocated wherein Mars and Earth share the same feeding zone close to Earth, and where Mars was subsequently scattered outwards to its current location. Based on dynamical modelling and isotopic analysis of Mars and major chondrite groups, Brasser et al. [2017] reach the opposite conclusion and argue instead that Mars' distinct isotopic composition calls for its formation in the asteroid belt with a subsequent dynamical pathway that parks it at its current orbit. The goals of the present study are to evaluate how the formation and migration of Jupiter could have influenced the nature of the material accreted by Mars, and to attempt to distinguish between these two scenarios. In what follows we rely on the output of dynamical simulations coupled with isotopic studies of primitive meteorites and samples originating from Mars (the SNC meteorites, which are thought to come from Mars) to constrain its accretion history.

Many primitive solar system objects are isotopically anomalous, meaning that their compositions cannot be related to the terrestrial isotopic composition by the laws of mass-dependent fractionation [Dauphas and Schauble, 2016 and references therein]. These isotopic anomalies – expressed in $\Delta^{17}O$, $\varepsilon^{48}Ca$, $\varepsilon^{50}Ti$, $\varepsilon^{54}Cr$, $\varepsilon^{64}Ni$, $\varepsilon^{92}Mo$, $\varepsilon^{100}Ru$, and $\varepsilon^{142}Nd$ notations – help establish links between planetary bodies, evaluate in quantitative ways the timing of delivery of various meteoritic building blocks to the terrestrial planets, and trace a planet's accretion through time [Dauphas, 2017]. Specifically, lithophile elements such as O, Ca, Ti and Nd trace the whole accretion history, whereas moderately siderophile Cr, Ni and Mo trace the latter half, and highly siderophile Ru only traces the last stages of planetary accretion.

# 2. Mars' accretion from N-body simulations

In Brasser et al. [2017], we tracked the fate of planetesimals and protoplanets in the framework of the Grand Tack model of terrestrial planet formation [Walsh et al., 2011] (see **Figure 1**). This scenario involves the inward-then-outward migration of Jupiter and is the only holistic model so far that can successfully account for the mass-semimajor axis distribution of the terrestrial planets, their accretion timescales, and the compositional structure of the asteroid belt [DeMeo & Carry, 2014].

The initial conditions of our N-body simulations are discussed in Brasser et al. [2016] and consist of a disk of planetary embryos and planetesimals between 0.7 and 3 AU, with a surface density profile that scales with heliocentric distance as $\Sigma \propto r^{-3/2}$; additional planetesimals that are assumed to have compositions similar to carbonaceous chondrites are initially placed between 5 and 9 AU. The initial conditions for these simulations follow the prescriptions of Jacobson and Morbidelli [2014]: the planetary embryos are of equal mass, being either 0.025, 0.05 or 0.08 Earth masses ($M_E$) and the total mass in planetesimals is either 1, 1/4 or 1/8 times the total mass in planetary embryos. The average total mass in the solid disk is 5 $M_E$. The permutation of these parameters results in nine sets of simulations, although here we only analyze simulation results using the latter two embryo masses because the outcome of the first has a low compatibility with the current solar system [Brasser et al., 2016]. When we assume that the embryos initially accrete from 10 km-sized planetesimals [Chambers, 2006], the time to grow to 0.05 and 0.08 $M_E$ takes between 0.5 to 1 Myr at 1 AU and 1 to 2 Myr at 1.5 AU [Chambers, 2006; Brasser et al., 2016]. Thus the solar system is assumed to be about 1 to 2 Myr old when the simulations are initiated. To begin, Jupiter and Saturn are placed at 3.5 and 4.5 AU with their current masses. The gas giants then migrate to 1.5 and 2.5 AU in 0.1 Myr from the start of the simulations and then move back to 5 AU and 7.5 AU over the next 5 Myr. Most of the migration is completed within 3 Myr from the start of the simulations, when the solar system is about 4-5 Myr old and the solar nebula has dissipated. Given the short duration of the inward migration in the simulation, the planetary embryos do not accrete much additional material before Jupiter tacks (see **Figure 1**), but because the embryos start with masses that are a significant fraction of Mars' final mass (typically 50-80%), their compositions to some extent still reflect the structure of the disk prior to Jupiter migration. The simulations were run for 150 Myr and perfect accretion is assumed [Brasser et al., 2016].

It has been suggested – but not universally accepted – that the solid disk originally had a heliocentric composition gradient consisting of dry, reduced enstatite chondrite (EC) close to the Sun, moderately volatile-rich and oxidised ordinary chondrite (OC) in the region of the asteroid belt, and highly oxidised and volatile-rich carbonaceous chondrites (CC) beyond [Morbidelli et al, 2012], mirroring the heliocentric distribution of asteroid groups [Gradie and Tedesco, 1982]. We compute the fractions of EC, OC and CC material that are accreted by the terrestrial planets by tracing the dynamical evolution of each protoplanet and planetesimal that are assembled into planets. Before the start of the simulation, planetary embryos are assumed to have coalesced locally so that their composition is either EC or OC depending on their initial distance to the Sun. We follow Brasser et al. (2017) and assume that EC are initially within 1.5 AU, OC are between 1.5 AU and 3 AU, and CC are assumed to have formed beyond Jupiter between 5 and 9 AU. Jupiter and Saturn initially reside at 3.5 and 4.5 AU. The final isotopic compositions of Earth and Mars sensitively depend on the location of the EC/OC boundary, with the best fit being in the vicinity of 1.5 AU.

At the beginning of the simulation, the typical mass of EC is 2.2 $M_E$ while that of OC is 3.0 $M_E$. The migration of Jupiter leads to the loss of on average 0.8 $M_E$ (36%) of EC and 2.7 $M_E$ (90%) of OC through ejection by, and collision with, the gas giants. In the aftermath of Jupiter's migration the truncated inner disk (within ~1 AU) is relatively well mixed and comprises, on average, 85% EC+14% OC+1% CC (compared to 100% EC in the same region before Jupiter's migration).

Dauphas [2017] divided the accretion of Earth into 3 stages (I, II, and III) corresponding to the first 60%, 60%-99.5% and the last 0.5% of Earth's accretion. In the dynamical simulations, we approximate Earth's stage I to be the first 5 Myr of evolution because analysis of the simulation data shows that by this time Venus and Earth reached roughly half their current masses (see also Jacobson & Walsh [2015]). Stages II and III combined trace the subsequent 145 Myr of the simulation. Roughly 80% of the mass that went into Earth consists of planetary embryos, ~95% of which are EC. The composition of material accreted by the Earth during stage I is 91% EC+8% OC+1% CC. For stages II and III, the accreted material is made of 85% EC+14% OC+1% CC, with relative uncertainties of about 1/5 of the listed values. This indicates a small decrease in the EC fraction with time, though it remains the main constituent of the Earth. Although Jupiter's migration preferentially depletes the OC reservoir, it also pushes some into the inner part of the disk, thus bringing more OC material to the growing Earth which therefore slightly decreases the proportion of EC. The OC material that is emplaced in the inner disk consists of both planetesimals and planetary embryos in the ratio set by the initial conditions.

By the time the giant planets have ceased migrating, the Mars analogues in our simulations consist (on average) of 67% EC+32% OC, but the 2σ uncertainties sample the range from 0% to 100% EC because the distribution of the accreted fraction of EC is bimodal [Brasser et al., 2017]: the mars analogues are predominantly mostly EC or OC but seldom a mixture of both. In our dynamical simulations, accretion onto Mars is highly stochastic and mostly consists of planetesimals with an average composition of 62% EC+38% OC, with relative uncertainties that are about 1/3 of the listed values.

At the end of the simulations, the Earth has a final composition of $87\%^{+9}_{-17}$ EC+$12\%^{+14}_{-10}$ OC, and Mars consists of $67\%^{+29}_{-66}$ EC+$32\%^{+65}_{-32}$ OC, with about 1% CC for each planet. Our composition estimates are robust: when the dividing distance between the EC and OC reservoirs is varied from 1 AU to 2 AU, the mean fraction of EC in both planets increases by less than 1/5 [Woo et al., 2018]. The large uncertainties in Mars' composition are a consequence of i) it accreting very little material during the simulations (on average 10%-50% of the final mass), ii) the bimodal distribution in the final fraction of EC in the Mars analogues, and iii) it undergoing a greater diversity in dynamical paths than Earth [Brasser et al, 2017]. Usually, but not always, both planets predominantly accrete EC during the early stages and see an increase in the delivery of OC material during the gas-driven migration of Jupiter and Saturn as a result of mixing some OC material into the inner disk. The dynamical simulations thus predict that the formation and migration of Jupiter caused the terrestrial planets to finalize their assembly from a dominantly reduced and relatively dry reservoir of planetary building blocks gleaned from the inner solar system.

A feature of the simulations is that Jupiter's migration stirs the disk and ejects the far majority of OC material. This is in contrast to classical simulations of terrestrial planet formation wherein the giant planets are assumed to reside on their current orbits. In those simulations all the terrestrial planets accrete material from the entire protoplanetary disk [Raymond et al., 2006; Brasser et al., 2016], and the feeding zone of each planet increases with time. This is true of Grand Tack as well [Fischer et al., 2018], but the nature of material that is accreted late is different for both models. In the classical model late accretion (mostly the latter half) consists primarily of material that originated in the region of the asteroid belt (assumed to be OC) while in the Grand Tack model, the material accreted after the tack is primarily EC because the OC reservoir has been depleted by Jupiter's migration.

The dynamical simulations alone are insufficient to pinpoint precisely where Mars formed due to the large uncertainties in its final composition. To better understand Mars' accretion history and formation location we attempt to apply the isotopic mixing model of Dauphas [2017] to Mars.

## 3. Mars' composition as inferred from its isotopic anomalies

It has long been known that the martian meteorites display isotopic anomalies for oxygen and several other elements. For Mars, previously-published meteorite mixtures based on isotopic studies are 75% CC+25% OC [Anderson, 1972], 85% H+11 CV+4% CI [Lodders & Fegley, 1997], 45% EC+55% OC [Sanloup et al., 1999] and 55% angrite +9% CI+36% OC [Fitoussi et al., 2016], where H is the high-metal component of the OC group. Tang & Dauphas [2014] showed that the mixture of Sanloup et al. [1999] matches Mars' $\Delta^{17}$O, $\varepsilon^{50}$Ti, $\varepsilon^{54}$Cr, $\varepsilon^{62}$Ni and $\varepsilon^{92}$Mo values, while that of Lodders & Fegley [1997] does not (see their Figure 2). All the proposed mixtures are clearly distinct from that of the Earth.

Dauphas [2017] suggests that Earth's isotopic composition is well-reproduced using four distinct classes of primitive (chondrite) meteorites as its building blocks: some material isotopically similar to, but chemically distinct from, enstatite chondrites (EC-component), ordinary chondrites that comprise H, L and LL groups (OC-component), and the CI and CO+CV classes of carbonaceous chondrites (CC-component). Combining these with the suggested three stages of accretion, Dauphas [2017] finds that an overall mixture of 71% EC+24% OC+5% COCV, consistent both with terrestrial isotopic measurements and with earlier independent estimates [Lodders, 2000; Javoy et al., 2010].

Here we use the same building blocks for Mars. Its accretion history can similarly be disentangled by examining the isotopic compositions of elements that partitioned differently between the mantle and core. No Ru isotopic measurements are available for martian material, disallowing a direct application of the methods of Dauphas [2017], but we can rely on O, Ti, Cr, Ni and Mo. The timing of the delivery of siderophile elements in the mantle depends on their affinity for metal [Dauphas, 2017]. We use the apparent Mars core/mantle distribution coefficients [Righter & Chabot, 2011] to estimate the time when the siderophile elements were delivered to the martian mantle ($D_{O,Ti}=0$, $D_{Cr}=3$, $D_{Mo}=56$, and $D_{Ni}=170$). The conditions of core formation in Mars differed significantly from those in the Earth, so that the isotopic signatures of martian meteorites record different fractions of the accretion history of Mars compared to Earth.

## 4. Methodology: Calculation of isotopic anomalies

We compute the constituents of the Earth and Mars using a Markov Chain Monte Carlo approach . The final isotopic composition of the mantle of Earth or Mars for each element $k$ is given by (Dauphas, 2017)

$$\varepsilon_{mantle,k} = \sum_j X_{j,k} \frac{\sum_i f_{i,j} C_{i,k} \varepsilon_{i,k}}{\sum_i f_{i,j} C_{i,k}} \quad (1).$$

The isotopic composition of each chondritic reservoir $i$ is $\varepsilon_{i,k}$, weighted by the element concentrations $C_{i,k}$. The values $f_{i,j}$ are the mass fractions accreted from each reservoir $i$ during accretion stage $j$, with the constraint $\sum_j f_{i,j} = 1$. The mass fraction of each element $k$ that is delivered to the mantle during accretion stage *j* is $X_{j,k}$ with the constraint $\sum_{j} X_{j,k} = 1$.

As mentioned in Section 2 the accretion for both planets is assumed to proceed through three stages. For both planets stage I consists of the first 60% of its mass. Stage II traces the next 39.3% (Earth) or 39.2% (Mars) while stage III consists of the remaining 0.7% (Earth) or 0.8% (Mars). The contribution from the last stage corresponds to the mass of material delivered to the Earth after the completion of core formation; the same is assumed for Mars. The amount of this late veneer is calculated based on the abundance of highly-siderophile elements in their mantles [Kimura et al., 1974; Rubie et al., 2015; Day et al., 2016]. Four different chondrite groups are considered as model building blocks: the carbonaceous CI and CO+CV (CO and CV have sufficiently similar isotopic compositions that we consider them a single reservoir), the enstatites EL and EH are lumped into the single EC chondrite supergroup, and ordinary chondrites H, L and LL are merged as the single OC supergroup. The elemental concentrations, $C_{i,k}$, of each supergroup are taken from Wasson & Kallemeyn [1988]. The isotopic variations of Mars and the various chondrite groups are taken from Dauphas [2017].

We estimate the values of $X_{j,k}$ from the probability density function (PDF) of the delivery of each element [Dauphas, 2017]. We operate under the assumptions that the planet grows from impacts of infinitesimal mass, and that the core mass fraction, $\beta$, the equilibration of the impactor core with the target metal, $k_{ic}$, and the metal-silicate partitioning coefficient, $D$, remain constant during the accretion. The PDF is then given by

$$\phi(x) = (1+\kappa) x^{\kappa} \quad (2),$$

$$\text{with } \kappa = \frac{D \beta k_{ic}}{1-\beta} \quad (3),$$

where $x$ is the mass fraction accreted by the planet. For Earth we set $k_{ic}$=0.4 [Dauphas, 2017], while for Mars we use $k_{ic}$=1 because its accretion most probably involved smaller impactors that were more likely to equilibrate with its mantle [Dauphas & Pourmand 2011; Kobayashi & Tanaka, 2010; Kobayashi and Dauphas, 2013; Brasser & Mojzsis, 2017; Mezger et al., 2013]. The Earth's core mass fraction is 0.33 while for Mars it is approximately 0.25 [Rivoldini et al., 2011]. The partition coefficients $D$ for each element used here are given in Supplementary Table 1 for Earth [Siebert et al., 2011; Badro et al., 2015] and Mars [Righter & Chabot, 2011]. The $D$–values correspond to the present bulk core-bulk mantle distribution coefficient ($D$=[C]$_{\text{bulk core}}$/[C]$_{\text{bulk mantle}}$). The range of acceptable $k_{ic}$ values for Earth is 0.3-0.8 [Nimmo et al., 2010].

A thorough discussion of the underlying assumptions of this model, the uncertainties in $D$ and $k_{ic}$ and how they affect the final results is beyond the scope of this paper. Dauphas [2017] evaluated their impacts by comparing the PDFs calculated from the analytical model with the PDFs calculated from more realistic and sophisticated accretion models and the conclusion was that the assumptions made did not affect the PDFs much. For Mars, where the accretion actually proceeded by impacts of small bodies (as opposed to giant impacts for the Earth), we expect the predictions of the analytical model to be even more reliable.

We compute the contributions $X_j$ from the partition coefficients in Supplementary Table 1 and from the PDFs assuming that stage III consisted of chondritic material. Thus, the contribution to the bulk silicate Earth (BSE) or bulk silicate Mars (BSM) from element Z during stage III (subscript *lv*), is given by $X_{lv}$ = [Z/Ru]$_{\text{CI}}$/[Z/Ru]$_{\text{BSE/BSM}}$, where it is assumed that all of the Ru in Earth's mantle arrived during stage III; for the Earth, this may not be a fully valid assumption [Rubie et al., 2016]. The value of $X_{lv}$ depends on the bulk elemental abundances for Earth [McDonough & Sun, 1995] and Mars [Brandon et al., 2012; Taylor, 2013]. The contributions to stages I and II follow from the PDF: $X_j = (1-X_{lv})(x_b^{\kappa+1} - x_a^{\kappa+1})/x_{lv}$, where $x_a$ and $x_b$ are the boundaries at each stage, and $x_{lv}$=0.993 or 0.992 for Earth and Mars, respectively. The resulting accretion contributions for both planets are listed in Supplementary Tables 2 and 3.

The free parameters of the model are the fractions of enstatite material during each stage ($f_{E,I}$, $f_{E,II}$ and $f_{E,III}$) and the proportions of CI, CO+CV and OC. For simplicity the last three are kept constant during the accretion. The model has five free parameters and five (for Mars) or six (for Earth) isotopic compositions that need to be reproduced. We determine the best fit for each of the five free input variables by minimising the $\chi^2$. We compute $\chi^2$ for Earth as [Dauphas et al., 2014]:

$$\chi^2 = \sum_k \sum_j X_{j,k} \frac{\left(\sum_i f_{i,j} C_{i,j} \varepsilon_{i,k}\right)^2}{\sum_i \left(f_{i,j} C_{i,j} \sigma_{i,k}\right)^2} \quad (4),$$

where $i$ is the reservoir index (maximum 4), $j$ is the stage index (maximum 3) and $k$ is the element index (maximum 6 for Earth). Here $\sigma_{i,k}$ are the standard deviations of the uncertainties in the isotopic composition of element $k$ from reservoir $i$, which are computed from the uncertainties in the measurements. For Mars the isotopic anomalies are non-zero and we use the relationship,

$$\chi^2 = \sum_k \frac{\left(\varepsilon_{mantle,k} - \varepsilon_{SNC,k}\right)^2}{\sigma^2_{mantle,k} + \sigma^2_{SNC,k}} \quad (5),$$

where the values of $\varepsilon_{mantle,k}$ are the compositions calculated from equation (1) and $\varepsilon_{SNC,k}$ are the measured isotopic compositions of the martian meteorites; the same applies to the uncertainties. The six isotopes to be computed are $\Delta^{17}O$, $\varepsilon^{50}Ti$, $\varepsilon^{54}Cr$, $\varepsilon^{64}Ni$, $\varepsilon^{92}Mo$ and $\varepsilon^{100}Ru$. Unlike for Earth, no isotopic data for Ru are available of Mars. $\varepsilon^{64}Ni$ has also not been measured precisely so we rely on existing $\varepsilon^{62}Ni$ isotope data for Mars and make use of the correlation between the nickel isotopes: $\varepsilon^{64}Ni = 2.8\varepsilon^{62}Ni$ [Tang & Dauphas, 2014]. Mars' stage III cannot be well traced and we only list our results without discussion.

We compute the uncertainties in the best-fit parameters using the Metropolis-Hastings MCMC algorithm [Metropolis et al., 1953; Hastings, 1970] wherein the accreted fractions were sampled using jumping normal distributions with widths of 0.03. The MCMC was run until the uncertainties converged (typically after a few million iterations). The MCMC calculation was implemented independently by two of us using different softwares (RB: Fortran; ND: Mathematica) and there is complete agreement between the results.

## 5. Results

We first applied our MCMC code to Earth. Our best-fit mixture during stage I is $51\%^{+28}_{-20}$ EC+$40\%^{+15}_{-24}$ OC+$0\%^{+5}_{-0}$ CI+$9\%^{+2}_{-5}$ COCV by mass. Stages II and III are dominated by EC: $100\%^{+0}_{-13}$ for stage II and $100\%^{+0}_{-6}$ for stage III. Uncertainties are $2\sigma$ and result from the uncertainties in isotopic data of the various chondrite groups. Stages I and II are different at greater than 95% confidence. Taken together, the best-fit meteorite mixture predicts that the Earth consists predominantly of EC [Javoy et al., 2010; Lodders, 2000]: $71\%^{+14}_{-12}$ EC+$24\%^{+10}_{-12}$ OC+$0\%^{+3}_{-0}$ CI+$5\%^{+1}_{-3}$ COCV ($\chi^2 \sim 11$). These values are identical to those of Dauphas [2017]. The high fractions of OC and COCV are needed to reproduce the terrestrial $\varepsilon^{50}Ti$ value, but given the uncertainties in the model end-member definitions (for example, EH and EL have slightly distinct $\varepsilon^{50}Ti$ values), it is conceivable that the contribution of OC was lower and that Earth accreted almost exclusively ($\geq 95\%$) from the EC component [Lodders, 2000; Javoy et al., 2010; Warren, 2011; Dauphas, 2017]. Some dynamical models of terrestrial planet formation not involving Jupiter's migration show that the last stages of the accretion of the terrestrial planets predominantly consist of material that originated in the region of the asteroid belt [Raymond et al., 2006; Fischer et al., 2018], assumed to be OC, while we show in Section 2 that when Jupiter migration is considered, the later stages of accretion are mostly EC. The conclusion reached based on isotopic anomalies differs from the prediction from the Grand Tack simulations: the latter cannot produce accretion made of 100% EC during the later stages.

For Mars, we report a best-fit composition of $68\%^{+0}_{-39}$ EC+$32\%^{+35}_{-0}$ OC with up to 3% in both CI and COCV ($\chi^2 \sim 12$). This compares favourably with the previous estimate of 45% EC+55% OC of Sanloup et al. [1999] but not with the 85% H+11% CV+4% CI of Lodders & Fegley [1997]. Note that these previous estimates relied primarily on $\Delta^{17}O$ while the present study uses the full array of isotopic anomalies measured in martian meteorites that became available since then. Mars' best-fit computed isotopic composition is $\Delta^{17}O = +0.32‰^{+0.31}_{-0.10}$, $\varepsilon^{50}Ti = -0.34^{+0.24}_{-0.08}$, $\varepsilon^{54}Cr = -0.18^{+0.09}_{-0.04}$, $\varepsilon^{64}Ni = -0.11^{+0.08}_{-0.02}$ and $\varepsilon^{92}Mo = +0.68^{+0.21}_{-0.17}$. The measured values are $\Delta^{17}O = +0.27‰ \pm 0.03‰$, $\varepsilon^{50}Ti = -0.54 \pm 0.17$, $\varepsilon^{54}Cr = -0.19 \pm 0.04$, $\varepsilon^{64}Ni = +0.10 \pm 0.28$ and $\varepsilon^{92}Mo = +0.20 \pm 0.53$ ($2\sigma$). The best-fit accretion mixture during stage I consists of $100\%^{+0}_{-89}$ EC+$0\%^{+85}_{-0}$ OC. Stage II contains a higher proportion of OC: $20\%^{+50}_{-17}$ EC+$80\%^{+11}_{-52}$ OC, while for stage III the fractions are $100\%^{+0}_{-95}$ EC+$0\%^{+90}_{-0}$ OC. All three stages contain up to 4% in both CI and COCV. A summary of the best-fit accretion histories of both Earth and Mars is shown in **Figure 2**. The uncertainties in the reconstructed accretion history of Mars are very large, both in the mixing model and in the dynamical model.

# 6. Discussion and conclusions

Mars' accretion took place in the first ~4 Myr of the solar system history. The early stages of its growth presumably took place before Jupiter migrated and the material accreted by Mars during that stage was primarily sourced locally. Both the dynamical model and the isotopic data favour accretion in the inner solar system and thus Mars' initial growth comprised a large fraction of EC. When Jupiter migrated in the first ~5 Myr of the solar system's history, it dynamically excited the disk. The inner disk was extensively stirred, thus obliterating the strong heliocentric gradient in isotopic compositions that existed before. Most OC material from the outer disk was ejected out of the solar system while some was mixed with EC; the latter therefore continued to dominate the inner solar system composition. The later stages of Mars' formation, posterior to Jupiter's migration, thus consisted of an EC-rich mixture. Jupiter's inward migration shepherded some OC material in its interior mean-motion resonances and left it in the inner disk, some of which was accreted by the growing terrestrial planets as the resonances swept past. The Earth, on the other hand, accreted primarily from embryos, many of which were formed in the pre-Jupiter migration EC-rich inner part of the inner disk. This sequence of events is consistent with the findings that Earth is largely dominated by EC material [Dauphas, 2017; this study] while Mars comprises a significant fraction of OC that was accreted most likely throughout Mars' formation history.

The uncertainties in the isotopic accretion model are very large (**Fig. 2**). **Figure 3** shows the fraction of EC accreted during each stage from the MCMC run. For all three stages the EC fraction is close to being uniformly random from 0 to 1. Thus both the dynamical simulations and the MCMC model are equally consistent with the scenarios proposed by Brasser et al. [2017] and Fitoussi et al. [2016]. In the former Mars forms in the asteroid belt and is dominated by OC material. In the latter Mars forms in the more internal regions that are dominated by EC material.

We close by adding a note of caution to the above results: there is *no consensus* on whether or not the Earth can be built from primitive meteorites *sensu stricto*. Achondrites plot outside of the range of isotopic compositions defined by chondrites for several elements [Warren 2011; Dauphas and Schauble 2016] and including them in the model may somewhat change the conclusions drawn here. In particular, angrites have oxygen isotopic compositions close to Earth and have been advocated as potential building blocks for the Earth and possibly Mars [Fitoussi et al., 2016]. They, however, have Ti, Cr and Ni isotopic compositions that are very distinct from the Earth and it takes fine tuning to reproduce the terrestrial composition for these elements, while this is readily explained by considering material akin to enstatite meteorites, which have near-terrestrial compositions for all isotopic systems measured thus far [Warren 2011; Dauphas and Schauble 2016; Dauphas 2017]. Furthermore, the absence of bulk concentration and isotopic data for Mo and Ru in most achondritic meteorites precludes us from incorporating them into our model. These elements provide critical constraints on Earth's accretion because the Earth is an end member in Ru and Mo isotopes [Dauphas et al. 2004; Fischer-Gödde et al., 2015]. For Mars, there are no data on Ru and the Mo isotopic data are of poor precision so most of the constraints on the late stage of Mars accretion are provided by Cr and Ni. More precise measurements of the isotopic compositions of Ni and Mo as well as Ru in martian samples would greatly improve our understanding of Mars' accretion history and distinguish between two end member scenarios for Mars' formation.

(4500 words; 0 remaining).

## Acknowledgements

The authors thank George Helffrich for stimulating discussions on metal-silicate partitioning, and Soko Matsumura and Jason Woo for their aid in analysing the dynamical simulations. We thank Alessandro Morbidelli and an anonymous reviewer for constructive criticism that helped improve this work. RB is grateful for financial support from JSPS KAKENHI (JP16K17662). ND is supported by grants from NSF (CSEDI EAR1502591 and Petrology and Geochemistry grant EAR1444951) and NASA (LARS NNX17AE86G, EW NNX17AE87G, and SSW NNX15AJ25G). RB and SJM acknowledge the Collaborative for Research in Origins (CRiO), which is supported by The John Templeton Foundation – FfAME Origins program: the opinions expressed in this publication are those of the authors, and do not necessarily reflect the views of the John Templeton Foundation. The source codes for the model used in this study are archived at the Earth Life Science Institute of the Tokyo Institute of Technology and the University of Chicago. The data, input and output files necessary to reproduce the figures are available from the authors upon request.

# Figures and captions

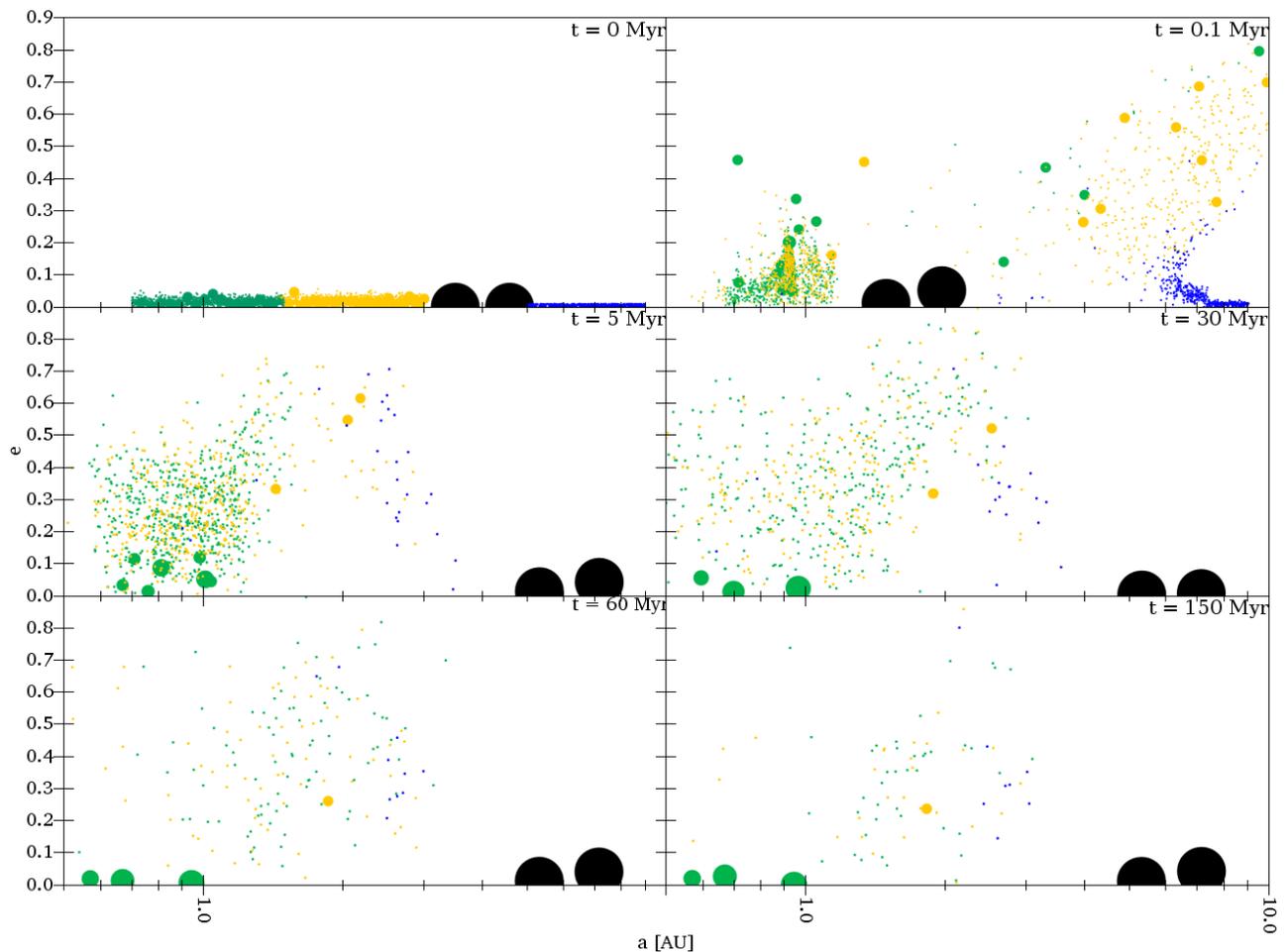

**Figure 1**. Snapshots of the evolution of the coupled migrations of Jupiter and Saturn in the Grand Tack model [Walsh et al., 2011]. The colour coding indicates the composition of planetesimals in the protoplanetary disk: green for EC, orange for OC and blue for CC. The gas giants Jupiter and Saturn migrate inwards and reach 1.5 AU in 0.1 Myr. They subsequently change directions (tack) and terminate their migration at their current positions after 5 Myr. During their journey, they remove >95% of the planetesimals and planetary embryos in the asteroid belt. Notably, they deplete the inner disk in OC material, leaving the terrestrial planets to grow from predominantly EC material in the later stages of their accretion.

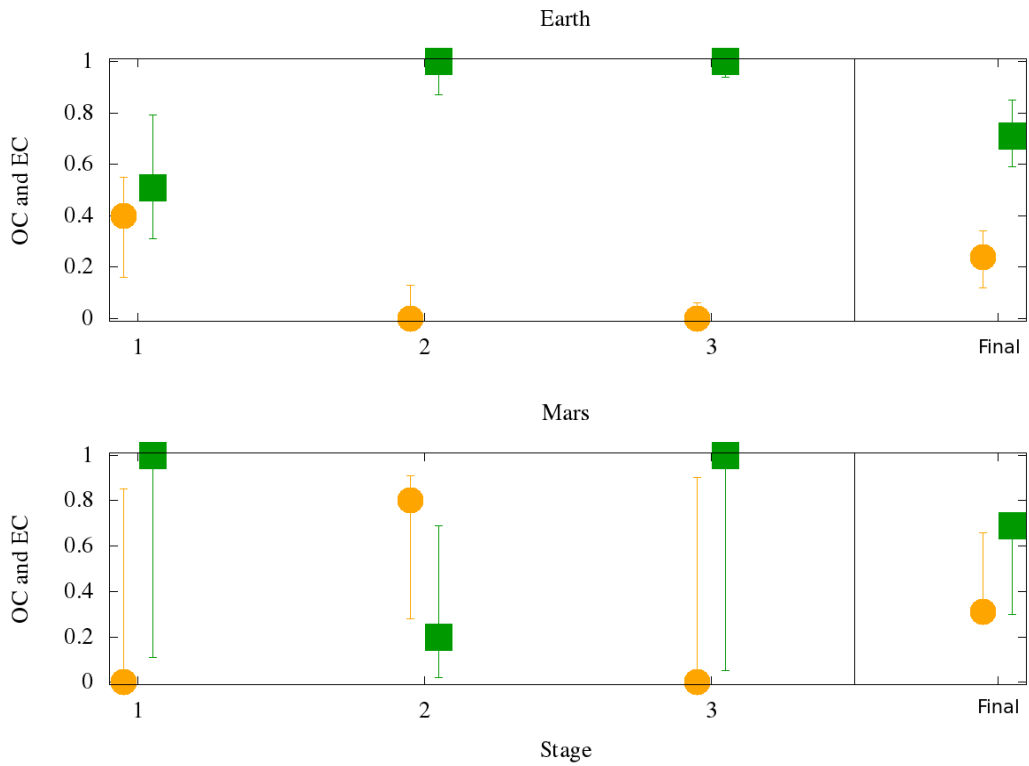

**Figure 2**. Proportions of the EC and OC component that go into Earth and Mars at each stage of their accretion calculated based on isotopic anomaly mixing models. Circles are OC, squares are EC. 'Final' is the overall composition of the planet. Since the CI and COCV reservoirs only contribute up to a few percent by mass to both planets, the EC + OC fraction is often close to 1.

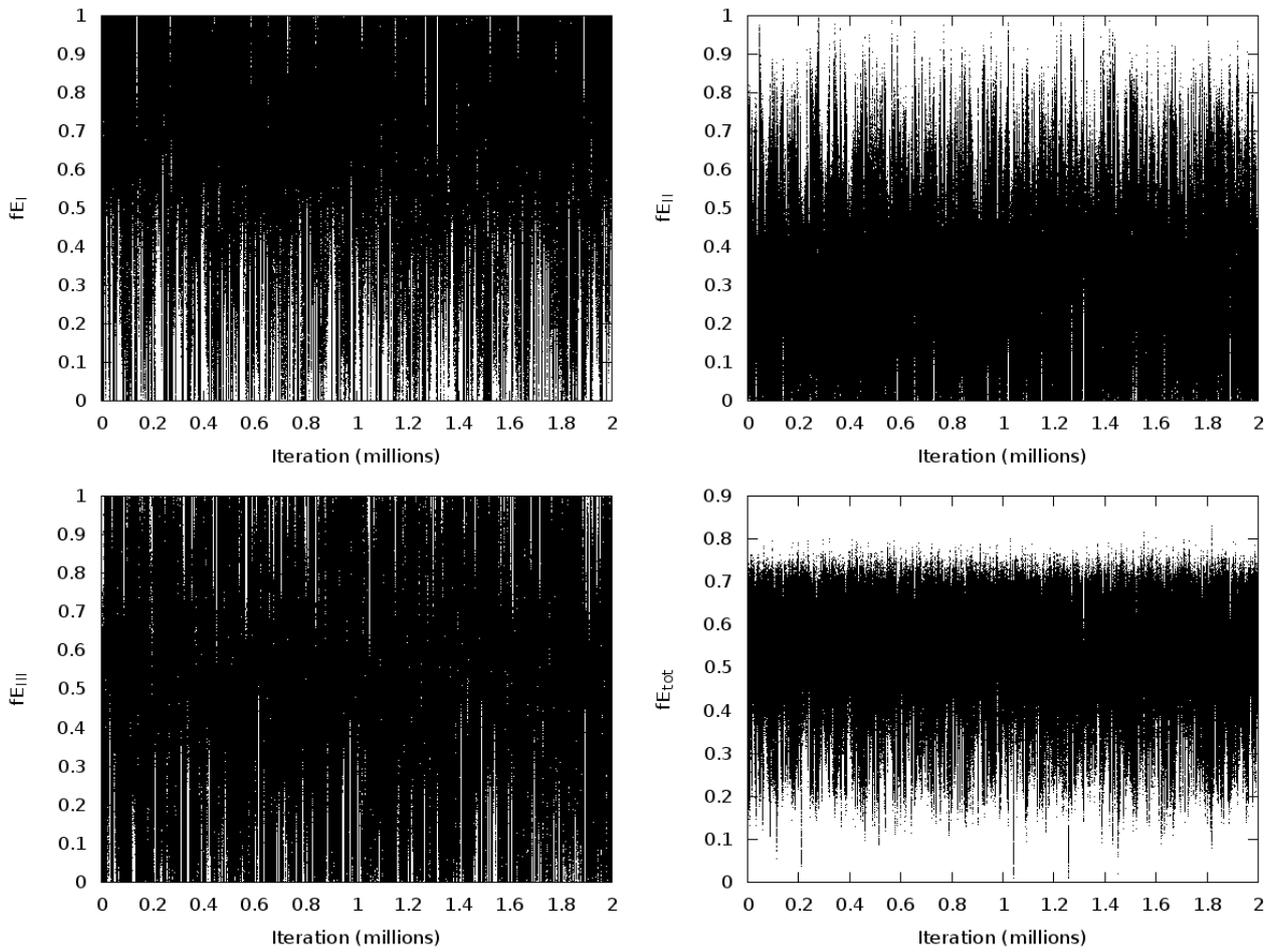

**Figure 3**. Fractions of EC accreted by Mars during stages I (top-left), II (top-right) and III (bottom-left) during our MCMC simulations with the Metropolis-Hastings algorithm. The bottom-right panel shows the total fraction of EC contained in Mars. It is obvious that for all three stages the fraction of EC is close to being uniformly random from 0 to 1, meaning that the low precision isotopic data available for SNC meteorites provide limited constraints on Mars' accretion history.

Supplementary tables

Table 1: Effective core/mantle partition coefficients, $D$ [Righter & Chabot, 2011; Siebert et al., 2011; Badro et al., 2015; Rai & Westrenen, 2013].

| Body/Element | Cr | Ni | Mo | Ru |
|---|---|---|---|---|
| Earth | 3 (2-4) | 26 (24-28) | 120 (90-150) | 1200 |
| Mars | 3 (<4.7) | 170 (155-200) | 56 (12-100) | 1600 (1000-1660) |

Table 2: Accretion fractional amounts for Earth

| stage | Begin | End | O | Ti | Cr | Ni | Mo | Ru |
|---|---|---|---|---|---|---|---|---|
| I | 0 | 0.6 | 0.6 | 0.6 | 0.443 | 0.042 | 0.0 | 0.0 |
| II | 0.6 | 0.993 | 0.393 | 0.393 | 0.549 | 0.919 | 0.873 | 0.0 |
| III | 0.993 | 1.0 | 0.007 | 0.007 | 0.007 | 0.038 | 0.127 | 1.0 |

Table 3: Accretion fractional amounts for Mars

| stage | Begin | End | O | Ti | Cr | Mo | Ni |
|---|---|---|---|---|---|---|---|
| I | 0 | 0.6 | 0.6 | 0.6 | 0.365 | 0.0 | 0.0 |
| II | 0.6 | 0.992 | 0.392 | 0.392 | 0.632 | 0.985 | 0.849 |
| III | 0.992 | 1.0 | 0.008 | 0.008 | 0.003 | 0.015 | 0.151 |